\documentclass[twocolumn, aps]{revtex4}%
\usepackage{amsfonts}
\usepackage{amsmath}
\usepackage{amssymb}
\usepackage{graphicx}%
\begin{document}
\title{Calculation of material properties and ray tracing in transformation media}
\author{D. Schurig$^{1}$, J.B. Pendry$^{2}$, D.R. Smith$^{1}$}
\affiliation{$^{1}$Duke University, Department of ECE, Box 90291, Durham, NC 27708}
\affiliation{$^{2}$Department of Physics, Blackett Laboratory, Imperial College London,
London SW7 2AZ, UK}

\begin{abstract}
Complex and interesting electromagnetic behavior can be found in spaces with
non-flat topology. When considering the properties of an electromagnetic
medium under an arbitrary coordinate transformation an alternative
interpretation presents itself. The transformed material property tensors may
be interpreted as a different set of material properties in a flat, Cartesian
space. We describe the calculation of these material properties for coordinate
transformations that describe spaces with spherical or cylindrical holes in
them. The resulting material properties can then implement invisibility cloaks
in flat space. We also describe a method for performing geometric ray tracing
in these materials which are both inhomogeneous and anisotropic in their
electric permittivity and magnetic permeability.

\end{abstract}
\date{31 May 2006}
\maketitle

\section{\bigskip Introduction}

Recently the use of coordinate transformations to produce material
specifications that control electromagnetic fields in interesting and useful
ways has been discussed. We have described such a method in which the
transformation properties of Maxwell's equations and the constitutive
relations can yield material descriptions that implement surprising
functionality, such as invisibility\cite{controlEM}. Another author described
a similar method where the two dimensional Helmholtz equation is transformed
to produce similar effects in the geometric limit\cite{ulfConformal}.\ We note
that these theoretical design methods are of more than academic interest, as
the material specifications can be implemented with metamaterial
technology\cite{mm1,mm2,mm3,terahertz,100terahertz,elcr}. There has also been
recent work on invisibility cloaking that does not employ the transformation
method\cite{naderCloak,miltonCloak}.

In this article we describe how to calculate these material properties
directly as Cartesian tensors and to perform ray tracing on devices composed
of these materials. We work out two examples in detail, the spherical and
cylindrical invisibility cloak.

\section{Material Properties}

We will describe the transformation properties of the electromagnetic,
material property tensors. We first note that the Minkowski form of Maxwell's
equations\cite{post,jackson,kongBook}
\begin{subequations}
\label{eq mink max}%
\begin{align}
F_{\alpha\beta,\mu}+F_{\beta\mu,\alpha}+F_{\mu\alpha,\beta}  &  =0\\
G_{,\alpha}^{\alpha\beta}  &  =J^{\beta}%
\end{align}
is form invariant for general space-time transformations. $F_{\alpha\beta}$ is
the tensor of electric field and magnetic induction, and $G^{\alpha\beta}$ is
the tensor density of electric displacement and magnetic field, and $J^{\beta
}$\ is the source vecto. In SI units the components are%
\end{subequations}
\begin{subequations}
\begin{align}
\left(  F_{\alpha\beta}\right)   &  =\left(
\begin{array}
[c]{cccc}%
0 & -E_{1} & -E_{2} & -E_{3}\\
E_{1} & 0 & B_{3} & -B_{2}\\
E_{2} & -B_{3} & 0 & B_{1}\\
E_{3} & B_{2} & -B_{1} & 0
\end{array}
\right) \\
\left(  G^{\alpha\beta}\right)   &  =\left(
\begin{array}
[c]{cccc}%
0 & -D_{1} & -D_{2} & -D_{3}\\
D_{1} & 0 & H_{3} & -H_{2}\\
D_{2} & -H_{3} & 0 & H_{1}\\
D_{3} & H_{2} & -H_{1} & 0
\end{array}
\right) \\
\left(  J^{\beta}\right)   &  =\left(
\begin{array}
[c]{c}%
\rho\\
J_{1}\\
J_{2}\\
J_{3}%
\end{array}
\right)
\end{align}
All of the information regarding the topology of the space is contained in the
constitutive relations%
\end{subequations}
\begin{equation}
G^{\alpha\beta}=\frac{1}{2}C^{\alpha\beta\mu\nu}F_{\mu\nu}%
\end{equation}
where $C^{\alpha\beta\mu\nu}$\ is the constitutive tensor representing the
properties of the medium, including its permittivity, permeability and
bianisotropic properties. $C^{\alpha\beta\mu\nu}$ is a tensor density of
weight +1, so it transforms as\cite{post}%
\begin{equation}
C^{\alpha^{\prime}\beta^{\prime}\mu^{\prime}\nu^{\prime}}=\left\vert
\det\left(  \Lambda_{\;\alpha}^{\alpha^{\prime}}\right)  \right\vert
^{-1}\Lambda_{\;\alpha}^{\alpha^{\prime}}\Lambda_{\;\beta}^{\beta^{\prime}%
}\Lambda_{\;\mu}^{\mu^{\prime}}\Lambda_{\;\nu}^{\nu^{\prime}}C^{\alpha\beta
\mu\nu} \label{eq trans c}%
\end{equation}
written in terms of the Jacobian transformation matrix%
\begin{equation}
\Lambda_{\;\alpha}^{\alpha^{\prime}}=\frac{\partial x^{\alpha^{\prime}}%
}{\partial x^{\alpha}} \label{eq trans matrix}%
\end{equation}
which is just the derivative of the transformed coordinates with respect to
the original coordinates. If we restrict ourselves to transformations that are
time invariant, the permittivity and permeability are also tensors
individually. Specifically, they are tensor densities of weight +1, which
transform as\cite{shyroki2,post}
\begin{subequations}
\label{eq trans e m}%
\begin{align}
\varepsilon^{i^{\prime}j^{\prime}}  &  =\left\vert \det\left(  \Lambda
_{\;i}^{i^{\prime}}\right)  \right\vert ^{-1}\Lambda_{\;i}^{i^{\prime}}%
\Lambda_{\;j}^{j^{\prime}}\varepsilon^{ij}\\
\mu^{i^{\prime}j^{\prime}}  &  =\left\vert \det\left(  \Lambda_{\;i}%
^{i^{\prime}}\right)  \right\vert ^{-1}\Lambda_{\;i}^{i^{\prime}}\Lambda
_{\;j}^{j^{\prime}}\mu^{ij}%
\end{align}
where the roman indices run from 1 to 3, for the three spatial coordinates, as
is standard practice. Equations (\ref{eq trans e m}) are the primary tools for
the transformation design method when the base medium is not bianisotropic and
the desired device moves or change shape with speeds much less than that of
light, i.e. probably all devices of practical interest. These equations can be
shown to be exactly equivalent to the results derived by Ward and
Pendry\cite{pendryWard}.

If the original medium is isotropic, (\ref{eq trans e m}) can also be written
in terms of the metric\cite{shyroki2,ulfGR}%
\end{subequations}
\begin{align}
\varepsilon^{i^{\prime}j^{\prime}}  &  =\left\vert \det\left(  g^{i^{\prime
}j^{\prime}}\right)  \right\vert ^{-1/2}g^{i^{\prime}j^{\prime}}\varepsilon\\
\mu^{i^{\prime}j^{\prime}}  &  =\left\vert \det\left(  g^{i^{\prime}j^{\prime
}}\right)  \right\vert ^{-1/2}g^{i^{\prime}j^{\prime}}\mu
\end{align}
where the metric is given by%
\begin{equation}
g^{i^{\prime}j^{\prime}}=\Lambda_{\;k}^{i^{\prime}}\Lambda_{\;l}^{j^{\prime}%
}\delta^{kl}%
\end{equation}

Maxwell's equations, (\ref{eq mink max}), together with the medium specified
by (\ref{eq trans c}) or (\ref{eq trans e m}) describe a single
electromagnetic behavior, but this behavior can be interpreted in two ways.

One way, the traditional way, is that the material property tensors that
appear on the left and right hand sides of (\ref{eq trans c}) or
(\ref{eq trans e m}) represent \emph{the same} material properties, but in
different spaces. The components in the transformed space are different form
those in the original space, due to the topology of the transformation. We
will refer to this view as the topological interpretation.

An alternative interpretation, is that the material property tensors on the
left and right hand sides of (\ref{eq trans c}) or (\ref{eq trans e m})
represent \emph{different} material properties. Both sets of tensor components
are interpreted as components in a flat, Cartesian space. The form invariance
of Maxwell's equations insures that both interpretations lead to the same
electromagnetic behavior. We will refer to this view as the materials interpretation.

To design something of interest, one imagines a space with some desired
property, a hole for example. Then one constructs the coordinate
transformation of the space with this desired property. Using
(\ref{eq trans c}) or (\ref{eq trans e m}) one can then calculate a set of
material properties that will implement this interesting property of the
imagined space in our own boring, flat, Cartesian space.

\subsection{Spherical Cloak}%

\begin{figure}
[t]
\begin{center}
\includegraphics[
height=3.026in,
width=3.3304in
]%
{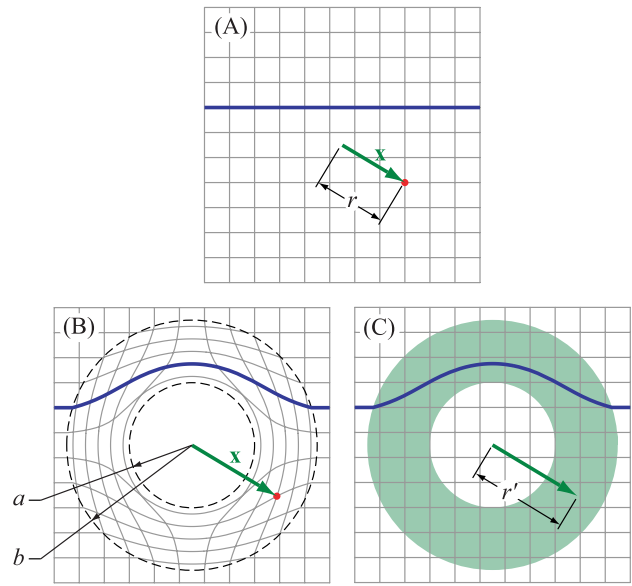}%
\caption{The thick blue line shows the path of the same ray in (A) the
original Cartesian space, and under two different interpretations of the
electromagnetic equations, (B) the topological interpretation and (C) the
materials interpretation. The position vector $\mathbf{x}$ is shown in both
the original and transformed spaces, and the length of the vector where the
transformed components are interpreted as Cartesian components is shown in
(C).}%
\label{fig transfig}%
\end{center}
\end{figure}
The spherical cloak is designed by considering a spherically symmetric
coordinate transformation. This transformation compresses all the space in a
volume of radius, $b$, into a spherical shell of inner radius, $a$, and outer
radius, $b$. Consider a position vector, $\mathbf{x}$. In the original
coordinate system (Fig.\ref{fig transfig}A) it has components, $x^{i}$, and in
the transformed coordinate system (Fig.\ref{fig transfig}B), $x^{i^{\prime}}$.
Of course, its magnitude, $r$, is independent of coordinate system
\begin{equation}
r=\left(  x^{i}x^{j}\delta_{ij}\right)  ^{1/2}=\left(  x^{i^{\prime}%
}x^{j^{\prime}}g_{i^{\prime}j^{\prime}}\right)  ^{1/2}%
\end{equation}
where $g_{i^{\prime}j^{\prime}}$\ is the metric of the transformed space. In
the materials interpretation, (Fig.\ref{fig transfig}C), we consider the
components, $x^{i^{\prime}}$, to be the components of a Cartesian vector, and
its magnitude is found using the appropriate flat space metric
\begin{equation}
r^{\prime}=\left(  x^{i^{\prime}}x^{j^{\prime}}\delta_{i^{\prime}j^{\prime}%
}\right)  ^{1/2}%
\end{equation}
Perhaps the simplest spherical cloak transformation maps points from a radius,
$r$, to a radius, $r^{\prime}$, according to the following linear function
\begin{equation}
r^{\prime}=\frac{b-a}{b}r+a\label{eq rp r}%
\end{equation}
which we apply over the domain, $0\leq r\leq b$, (or equivalently, $a\leq
r^{\prime}\leq b$). Outside this domain we assume an identity transformation.
(All equations in the remainder of this article apply only to the
transformation domain.) We must always limit the transformation to apply only
over a finite region of space if we wish to implement it with materials of
finite extent. Note that when $r=0$ then $r^{\prime}=a$, so that the origin is
mapped out to a finite radius, opening up a whole in space. Note also that
when $r=b$ then $r^{\prime}=b$, so that space at the outer boundary of the
transformation is undistorted and there is no discontinuity with the space
outside the transformation domain.

Now since our transformation is radially symmetric, the unit vectors in
materials interpretation and in the original space must be equal.%
\begin{equation}
\frac{x^{i^{\prime}}}{r^{\prime}}=\frac{x^{i}}{r}\delta_{\;i}^{i^{\prime}}%
\end{equation}
Expressing the components of the position vector in the transformed space in
terms of only the components in the original space, using (\ref{eq rp r}), we
obtain.%
\begin{equation}
x^{i^{\prime}}=\frac{b-a}{b}x^{i}\delta_{\;i}^{i^{\prime}}+a\frac{x^{i}}%
{r}\delta_{\;i}^{i^{\prime}}%
\end{equation}
Now that we have this expression, we need not worry about the interpretations
of transformed space, we can just proceed in standard fashion to compute the
transformation matrix. To take the derivative of this expression we note that%
\begin{equation}
\frac{\partial}{\partial x^{j}}\frac{x^{i}}{r}=-\frac{x^{i}x^{k}\delta_{kj}%
}{r^{3}}+\frac{1}{r}\delta_{\;j}^{i}%
\end{equation}
and obtain the transformation matrix%
\begin{equation}
\Lambda_{\;j}^{i^{\prime}}=\frac{r^{\prime}}{r}\delta_{\;j}^{i^{\prime}}%
-\frac{ax^{i}x^{k}\delta_{\;i}^{i^{\prime}}\delta_{kj}}{r^{3}}%
\end{equation}
The components of this expression written out are%
\begin{equation}
\left(  \Lambda_{\;j}^{i^{\prime}}\right)  =\left(
\begin{array}
[c]{ccc}%
\frac{r^{\prime}}{r}-\frac{ax^{2}}{r^{3}} & -\frac{axy}{r^{3}} & -\frac
{axz}{r^{3}}\\
-\frac{ayx}{r^{3}} & \frac{r^{\prime}}{r}-\frac{ay^{2}}{r^{3}} & -\frac
{ayz}{r^{3}}\\
-\frac{azx}{r^{3}} & -\frac{azy}{r^{3}} & \frac{r^{\prime}}{r}-\frac{az^{2}%
}{r^{3}}%
\end{array}
\right)
\end{equation}
To calculate the determinant of this matrix we note that we can always rotate
into a coordinate system such that%
\begin{equation}
\left(  x^{i}\right)  =(r,0,0)
\end{equation}
then the determinant is, by inspection, given by%
\begin{equation}
\det(\Lambda_{\;j}^{i^{\prime}})=\frac{r^{\prime}-a}{r}\left(  \frac
{r^{\prime}}{r}\right)  ^{2}%
\end{equation}
If we assume that our original medium is free space, then the permittivity and
permeability will be equal to each other. As a short hand, we define a tensor,
$n^{i^{\prime}j^{\prime}}$, to represent both.%
\begin{equation}
n^{i^{\prime}j^{\prime}}\equiv\varepsilon^{i^{\prime}j^{\prime}}%
=\mu^{i^{\prime}j^{\prime}}%
\end{equation}
Though this definition is suggestive of refractive index, $n^{i^{\prime
}j^{\prime}}$ would only represent the scalar refractive index if the
permittivity and permeability were additionally isotropic, which is not the
case here.

Working out the algebra, we find that the material properties are then given
by%
\begin{equation}
n^{i^{\prime}j^{\prime}}=\frac{b}{b-a}\left[  \delta^{i^{\prime}j^{\prime}%
}-\frac{2ar^{\prime}-a^{2}}{r^{\prime4}}x^{i^{\prime}}x^{j^{\prime}}\right]
\end{equation}
where we have eliminated any dependence on the components of $\mathbf{x}$ in
the original space, $x^{i}$, or the magnitude, $r$. We can now drop the primes
for aesthetic reasons, and we need not make the distinction between vectors
and one-forms as we consider this to be a material specification in flat,
Cartesian, three-space, where such distinctions are not necessary. Writing
this expression in direct notation%
\begin{equation}
\mathbf{n}=\frac{b}{b-a}\left(  \mathbf{I}-\frac{2ar-a^{2}}{r^{4}}%
\mathbf{r}\otimes\mathbf{r}\right)  \label{eq spherical n}%
\end{equation}
where $\mathbf{r}\otimes\mathbf{r}$\ is the outer product of the position
vector with itself, also referred to as a dyad formed from the position
vector. We note, for later use, that the determinant can be easily calculated,
as above, using an appropriately chosen rotation%
\begin{equation}
\det\left(  \mathbf{n}\right)  =\left(  \frac{b}{b-a}\right)  ^{3}\left(
\frac{r-a}{r}\right)  ^{2} \label{eq shperical detn}%
\end{equation}

\subsection{\bigskip Cylindrical Cloak}

To analyze a cylindrical cloak we will use two projection operators. One which
projects onto the cylinder's axis, (which we will call the third coordinate or
$z$-axis), and one that projects onto the plane normal to the cylinder's axis.%
\begin{subequations}
\begin{align}
Z^{ij}  &  =\delta_{\;3}^{i}\delta_{\;3}^{j}\\
T^{ij}  &  =\delta_{\;1}^{i}\delta_{\;1}^{j}+\delta_{\;2}^{i}\delta_{\;2}^{j}%
\end{align}
We do not mean to imply that these are tensors. We define these operators to
perform these projections onto the third coordinate and the plane normal to
the third coordinate in whatever basis (including mixed bases) they are
applied to. Thus we will refer to their components with indices up or down,
primed or un-primed, at will. We now use the transverse projection operator to
define a transverse coordinate.%
\end{subequations}
\begin{equation}
\rho^{i}=T_{\;j}^{i}x^{j}%
\end{equation}
The coordinate transformation for the cylindrical case is the same as that of
the spherical case in the two dimensions normal to the cylinder's axis. Along
the cylinders axis the transformation is the identity. Thus we have for the
transformation matrix.%
\begin{equation}
\left(  \Lambda_{\;j}^{i^{\prime}}\right)  =\left(
\begin{array}
[c]{ccc}%
\frac{\rho^{\prime}}{\rho}-\frac{ax^{2}}{\rho^{3}} & -\frac{axy}{\rho^{3}} &
0\\
-\frac{ayx}{\rho^{3}} & \frac{\rho^{\prime}}{\rho}-\frac{ay^{2}}{\rho^{3}} &
0\\
0 & 0 & 1
\end{array}
\right)
\end{equation}
or written in index form%
\begin{equation}
\Lambda_{\;j}^{i^{\prime}}=\frac{\rho^{\prime}}{\rho}T_{\;j}^{i^{\prime}%
}-\frac{a\rho^{i}\rho^{k}\delta_{\;i}^{i^{\prime}}\delta_{kj}}{\rho^{3}%
}+Z_{\;j}^{i^{\prime}} \label{eq cyl trans}%
\end{equation}
Again, we can easily calculate the determinant by rotating into a coordinate
system where%
\begin{equation}
\left(  x^{i}\right)  =\left(  \rho^{i}\right)  =(\rho,0,0)
\end{equation}
then we find the determinant to be%
\begin{equation}
\det(\Lambda_{\;j}^{i^{\prime}})=\frac{\rho^{\prime}-a}{\rho}\frac
{\rho^{\prime}}{\rho}%
\end{equation}
The material properties in direct notation and dropping the primes are%
\begin{equation}
\mathbf{n}=\frac{\rho}{\rho-a}\mathbf{T}-\frac{2a\rho-a^{2}}{\rho^{3}\left(
\rho-a\right)  }\boldsymbol{\rho}\otimes\boldsymbol{\rho}+\left(  \frac
{b}{b-a}\right)  ^{2}\frac{\rho-a}{\rho}\mathbf{Z} \label{eq cylindrical n}%
\end{equation}
Again we note the determinant for later use, which takes the rather simple
form%
\begin{equation}
\det\left(  \mathbf{n}\right)  =\left(  \frac{b}{b-a}\right)  ^{2}\frac
{\rho-a}{\rho} \label{eq cylindrical detn}%
\end{equation}

\section{Hamiltonian and Ray Equations}

The Hamiltonian we will use for generating the ray paths is essentially the
plane wave dispersion relation\cite{inhomoGeoOptics}. We derive it here,
briefly, to show our choice of dimensionality for the relevant variables. We
begin with Maxwell's curl equations in SI\ units%
\begin{equation}
\mathbf{\nabla\times E}=-\frac{\partial\mathbf{B}}{\partial t}%
\;\;\;\;\;\;\;\;\mathbf{\nabla\times H}=\frac{\partial\mathbf{D}}{\partial t}
\label{eq curl 1}%
\end{equation}
We assume plane wave solutions with slowly varying coefficients, appropriate
for the geometric limit
\begin{equation}
\mathbf{E}=\mathbf{E}_{0}e^{i\left(  k_{0}\mathbf{k\cdot x}-\omega t\right)
}\;\;\;\;\;\;\;\;\mathbf{H}=\frac{1}{\eta_{0}}\mathbf{H}_{0}e^{i\left(
k_{0}\mathbf{k\cdot x}-\omega t\right)  } \label{eq plane waves}%
\end{equation}
Here $\eta_{0}=\sqrt{\mu_{0}/\varepsilon_{0}}$\ is the impedance of free
space, giving $\mathbf{E}_{0}$\ and $\mathbf{H}_{0}$\ the same units, and
$k_{0}=\omega/c$ making $\mathbf{k}$ dimensionless. We use constitutive
relations with dimensionless tensors $\boldsymbol{\varepsilon}$\ and
$\boldsymbol{\mu}$.
\begin{equation}
\mathbf{D}=\varepsilon_{0}\boldsymbol{\varepsilon}\mathbf{E\;\;\;\;\;\;\;\;B}%
=\mu_{0}\boldsymbol{\mu}\mathbf{H} \label{eq constituitive relations}%
\end{equation}
Plugging (\ref{eq plane waves}) and (\ref{eq constituitive relations})\ into
the curl equations (\ref{eq curl 1}) we obtain%
\begin{equation}
\mathbf{k}\times\mathbf{E}_{0}-\boldsymbol{\mu}\mathbf{H}_{0}%
=0\;\;\;\;\;\;\;\;\mathbf{k}\times\mathbf{H}_{0}+\boldsymbol{\varepsilon
}\mathbf{E}_{0}=0
\end{equation}
Eliminating the magnetic field we find
\begin{equation}
\mathbf{k}\times\left(  \boldsymbol{\mu}^{-1}\left(  \mathbf{k}\times
\mathbf{E}_{0}\right)  \right)  +\boldsymbol{\varepsilon}\mathbf{E}_{0}=0
\label{eq dispersion 1}%
\end{equation}
Defining the operator, $\mathbf{K}$\cite{kongBook},
\begin{equation}
K_{ik}\equiv\epsilon_{ijk}k_{j}%
\end{equation}
the dispersion relation (\ref{eq dispersion 1}) can be expressed as a single
operator on $\mathbf{E}_{0}$,
\begin{equation}
\left(  \mathbf{K}\boldsymbol{\mu}^{-1}\mathbf{K}+\boldsymbol{\varepsilon
}\right)  \mathbf{E}_{0}=0
\end{equation}
which must be singular for non-zero field solutions. The dispersion relation
expresses that this operator must have zero determinate.
\begin{equation}
\det\left(  \mathbf{K}\boldsymbol{\mu}^{-1}\mathbf{K}+\boldsymbol{\varepsilon
}\right)  =0
\end{equation}
Now for material properties derived from transforming free space,
$\boldsymbol{\varepsilon}$\ and $\boldsymbol{\mu}$\ are the same symmetric
tensor, which we\ call $\mathbf{n}$. In this case the dispersion relation has
an alternate expression%
\begin{equation}
\det\left(  \mathbf{Kn}^{-1}\mathbf{K}+\mathbf{n}\right)  =\frac{1}%
{\det\left(  \mathbf{n}\right)  }\left(  \mathbf{knk-}\det\left(
\mathbf{n}\right)  \right)  ^{2}%
\end{equation}
This can be proved by brute force, evaluating the two expression for a matrix,
$\mathbf{n}$, with arbitrary symmetric components, or perhaps some other
clever way. The latter expression is clearly fourth order in $\mathbf{k}$, but
has only two unique solutions. Thus we discover that in media with
$\boldsymbol{\varepsilon}=\boldsymbol{\mu}$ the ordinary ray and extraordinary
ray (found in anisotropic dielectrics) are degenerate. This can also be seen
by noting that in free space the ordinary ray and extraordinary ray follow the
same path. A coordinate transformation cannot separate these two paths, so
they will follow the same path in the transformed coordinate space and thus
also in the equivalent media.

The Hamiltonian then easily factors into two terms that represent degenerate
modes. Further it is easy to show ( by plugging (\ref{eq H})\ into
(\ref{eq motion})\ ) that the Hamiltonian may be multiplied by an arbitrary
function of the spatial coordinates without changing the paths obtained from
the equations of motion, (only the parameterization is changed), thus we can
drop the factor, $1/\det\left(  \mathbf{n}\right)  $, and our Hamiltonian is%

\begin{equation}
H=f\left(  \mathbf{x}\right)  \left(  \mathbf{knk}-\det\left(  \mathbf{n}%
\right)  \right)  \label{eq H}%
\end{equation}
where $f\left(  \mathbf{x}\right)  $ is some arbitrary function of position.
The equations of motion are\cite{inhomoGeoOptics}
\begin{subequations}
\label{eq motion}%
\begin{align}
\frac{d\mathbf{x}}{d\tau}  &  =\frac{\partial H}{\partial\mathbf{k}}\\
\frac{d\mathbf{k}}{d\tau}  &  =-\frac{\partial H}{\partial\mathbf{x}}%
\end{align}
where $\tau$ parameterizes the paths. This pair of coupled, first order,
ordinary differential equations can be integrated using a standard solver,
such as Mathematica's NDSolve.

\section{\bigskip Refraction}

The equations of motion, (\ref{eq motion}), govern the path of the ray
throughout the continuously varying inhomogenous media. At discontinuities,
such as the outer boundary of a cloak, we must perform boundary matching to
determine the discontinuous change in direction of the ray, i.e. refraction.
Given $\mathbf{k}_{1}$ on one side of the boundary we find $\mathbf{k}_{2}$ on
the other side as follows. The transverse component of the wave vector is
conserved across the boundary.
\end{subequations}
\begin{equation}
\left(  \mathbf{k}_{1}-\mathbf{k}_{2}\right)  \times\mathbf{n}=\mathbf{0}%
\end{equation}
where here $\mathbf{n}$ is the unit normal to the boundary. This vector
equation represents just two equations. The third is obtained by requiring the
wave vector to satisfy the plane wave dispersion relation of the mode
represented by the Hamiltonian.%
\begin{equation}
H\left(  \mathbf{k}_{2}\right)  =0
\end{equation}
These three equations determine the three unknowns of the vector components of
$\mathbf{k}_{2}$. Since $H$ is quadratic in $\mathbf{k}$, there will be two
solutions, one that carries energy into medium 2, the desired solution, and
one that carries energy out. The path of the ray, $d\mathbf{x}/d\tau$,
determines the direction of energy flow, so the Hamiltonian can be used to
determine which is the desired solution. The desired solution satisfies%
\begin{equation}
\frac{\partial H}{\partial\mathbf{k}}\cdot\mathbf{n}>0
\end{equation}
if $\mathbf{n}$ is the normal pointing into medium 2.

These equations apply equally well to refraction into or out of transformation
media. Refracting out into free space is much easier since the Hamiltonian of
free space is just, $H=\mathbf{k}\cdot\mathbf{k}-1$.

\section{Cloak Hamiltonians}

We now show specific examples of ray tracing. Below we will choose a specific
form for the Hamiltonian, plug in the material properties and display the
derivatives of the Hamiltonian, for both the spherical and cylindrical cloak.

\subsection{Spherical Cloak}%

\begin{figure}
[ptb]
\begin{center}
\includegraphics[
trim=0.865794in 1.154042in 0.962925in 0.385735in,
height=2.8193in,
width=2.5374in
]%
{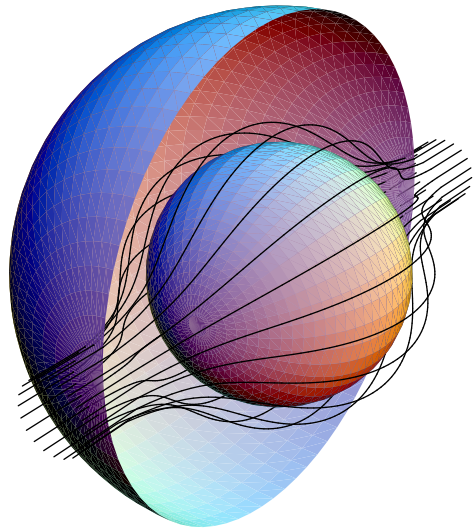}%
\caption{Rays traversing a spherical cloak. The transformation media that
comprises the cloak lies between the two spheres.}%
\label{fig sphere}%
\end{center}
\end{figure}
For the spherical cloak (Fig.\ref{fig sphere}), the Hamiltonian which yields
the simplest equations is%
\begin{equation}
H=\frac{1}{2}\frac{b-a}{b}\left(  \mathbf{knk}-\det\left(  \mathbf{n}\right)
\right)
\end{equation}
Plugging in the material properties from (\ref{eq spherical n}) and
(\ref{eq shperical detn}) we obtain%
\[
H=\frac{1}{2}\mathbf{k}\cdot\mathbf{k}-\frac{1}{2}\frac{2ar-a^{2}}{r^{4}%
}\left(  \mathbf{x}\cdot\mathbf{k}\right)  ^{2}-\frac{1}{2}\left[
\frac{b\left(  r-a\right)  }{r\left(  b-a\right)  }\right]  ^{2}%
\]
Taking the derivatives, (which is straight forward particularly in index
form), yields
\begin{subequations}
\begin{align}
\frac{\partial H}{\partial\mathbf{k}} &  =\mathbf{k}-\frac{2ar-a^{2}}{r^{4}%
}\left(  \mathbf{x}\cdot\mathbf{k}\right)  \mathbf{x}\\
\frac{\partial H}{\partial\mathbf{x}} &  =\left[
\begin{array}
[c]{c}%
-\frac{2ar-a^{2}}{r^{4}}\left(  \mathbf{x}\cdot\mathbf{k}\right)
\mathbf{k}+\frac{3ar-2a^{2}}{r^{6}}\left(  \mathbf{x}\cdot\mathbf{k}\right)
^{2}\mathbf{x}\\
-\left(  \frac{b}{b-a}\right)  ^{2}\left(  \frac{ar-a^{2}}{r^{4}}\right)
\mathbf{x}%
\end{array}
\right]
\end{align}

\subsection{Cylindrical Cloak}%

\begin{figure}
[ptb]
\begin{center}
\includegraphics[
height=1.8948in,
width=3.3399in
]%
{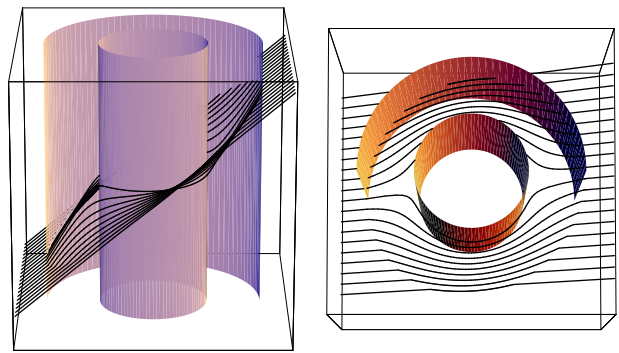}%
\caption{Rays traversing a cylindrical cloak at an oblique angle. The
transformation media that comprises the cloak lies in an annular region
between the cylinders.}%
\label{fig cylinder}%
\end{center}
\end{figure}
For the cylindrical cloak (Fig.\ref{fig cylinder}), the Hamiltonian which
yields the simplest equations is%

\end{subequations}
\begin{equation}
H=\frac{1}{2}\frac{\rho-a}{\rho}\left(  \mathbf{knk}-\det\left(
\mathbf{n}\right)  \right)
\end{equation}
Plugging in the material properties form (\ref{eq cylindrical n}) and
(\ref{eq cylindrical detn}) we obtain%
\begin{equation}
H=\frac{1}{2}\mathbf{kTk}-\frac{1}{2}\frac{2a\rho-a^{2}}{\rho^{4}}\left(
\boldsymbol{\rho}\cdot\mathbf{k}\right)  ^{2}+\frac{1}{2}\left[
\frac{b\left(  \rho-a\right)  }{\rho\left(  b-a\right)  }\right]  ^{2}\left(
\mathbf{kZk}-1\right)
\end{equation}
For taking the derivatives we note that the derivative of the transverse
position vector with respect to the position vector is the transverse
projection operator.%
\begin{equation}
\frac{\partial\boldsymbol{\rho}}{\partial\mathbf{x}}=\mathbf{T}%
\end{equation}
The derivatives are thus
\begin{subequations}
\begin{align}
\frac{\partial H}{\partial\mathbf{k}}  &  =\mathbf{Tk}-\frac{2a\rho-a^{2}%
}{\rho^{4}}\left(  \boldsymbol{\rho}\cdot\mathbf{k}\right)  \boldsymbol{\rho
}+\left[  \frac{b\left(  \rho-a\right)  }{\rho\left(  b-a\right)  }\right]
^{2}\mathbf{Zk}\\
\frac{\partial H}{\partial\mathbf{x}}  &  =\left[
\begin{array}
[c]{c}%
\frac{3a\rho-2a^{2}}{\rho^{6}}\left(  \boldsymbol{\rho}\cdot\mathbf{k}\right)
^{2}\boldsymbol{\rho}-\frac{2a\rho-a^{2}}{\rho^{4}}\left(  \mathbf{\rho\cdot
k}\right)  \mathbf{Tk}\\
+\left(  \frac{b}{b-a}\right)  ^{2}\frac{a\rho-a^{2}}{\rho^{4}}\left(
\mathbf{kZk}-1\right)  \boldsymbol{\rho}%
\end{array}
\right]
\end{align}

\section{Conclusion}

We have shown how to calculate the material properties associated with a
coordinate transformation and use these properties to perform ray tracing.
Examples, of spherical and cylindrical cloaks are worked out in some detail.
Some of the value in this effort is to provide independent confirmation that
the material properties calculated from the transformation do indeed cause
electromagnetic waves to behave in the desired and predicted manner.
Eventually, this technique will become more accepted and independent
confirmation will not be needed. One can see what the waves will do much more
easily by applying the transformation to the rays or fields in the original
space where the behavior is simpler if not trivial. However, one may still
want to perform ray tracing on these media to see the effects of perturbations
from the ideal material specification.

D. Schurig wishes to acknowledge the IC Postdoctoral Research Fellowship program.

\bibliographystyle{apsrev}
\bibliography{articles,dav}

\end{subequations}
\end{document}